\documentstyle[prl,aps,epsfig,amsmath]{revtex}
\begin{document}
\twocolumn[\hsize\textwidth\columnwidth\hsize\csname@twocolumnfalse\endcsname
\title  {Classical Langevin Dynamics for Model Hamiltonians}
\author{Morrel H. Cohen}
\address{Department of Physics and Astronomy\\
Rutgers University, Piscataway, NJ 08854-8019 USA}

\maketitle

\begin{abstract}
We propose a scheme for extending  the model Hamiltonian method
developed originally for studying the equilibrium properties of
complex perovskite systems to include Langevin dynamics.  The
extension is based on Zwanzig's treatment of nonlinear
generalized Langevin's equations.  The parameters entering the
equations of motion are to be determined by mapping from
first-principles calculations, as in the original model
Hamiltonian method.  The scheme makes possible, in principle, the
study of the dynamics and kinetics of structural transformations
inaccessible to the original model Hamiltonian method.  Moreover,
we show that the equilibrium properties are governed by an
effective Hamiltonian which differs from that used in previous
work by a term which captures the coherent part of the previously
ignored dynamical interaction with the omitted degrees of freedom.
We describe how the additional information required for the
Langevin equations can be obtained by a minor extension of the
previous mapping.\\
\end{abstract}


\narrowtext
]


This paper is dedicated to Professor  Josef Devreese on the
occasion of his 65th birthday and his formal retirement.  May he
continue to have many more productive years in theoretical
physics.
\\

\section{INTRODUCTION}
Our understanding of the properties of relativity simple
materials has advanced markedly over the last half century.  A
wealth of experimental data exists with which theory, both as
formal analysis and as first-principles computation, is in
quantitative agreement.  More recently, attention has turned to
such complex materials as multication oxides which, even at this
advanced stage of condensed matter physics and materials science,
still pose a compelling challenge.  Those materials possess an
immensely rich phenomenology - antiferromagnetism and
high-temperature super-conductivity in cuprates; colossal
magnetoresistance, charge, spin, and orbital ordering in manganates;
ferroelectricity, antiferroelectricity, ferrodistortion, and
relaxors in perovskites and related materials; and, more recently,
a colossal temperature-independent dielectric constant in single
crystal CaCu$_{3}$Ti$_{4}$O$_{12}$ (CCTO).  First-principles
studies of these complex oxides, however, have been limited to
equilibrium properties or responses of perfect crystals and to
relatively simple multilayer structures.  More difficult
equilibrium questions have been addressed for the perovskite
ferroelectrics by constructing a simplified model Hamiltonian the
parameters of which are obtained from first principles
computations.  The construction of the model Hamiltonians for the
perovskite ferroelectrics proceeds as follows. Those optical
branches which are unstable in the high symmetry cubic structure
are identified.$^{1}$  From them, local modes are constructed
which approximate the optimally localized lattice Wannier
functions$^{2}$ and which capture the dominant anharmonicities in
the total energy.  These local modes form the basis for the
construction of an ``effective'' or ``model'' Hamiltonian$^{2}$
simple enough to use for Monte Carlo simulations of equilibrium
structures and properties.$^{3-5}$  The parameters of the model
Hamiltonian are few enough in number to admit computation by
first-principles methods via a mapping to the model Hamiltonian.

There are many dynamic or kinetic phenomena of great interest
which cannot be addressed via the model Hamiltonian scheme at its
present level of development because the model Hamiltonians are
too incomplete to use for molecular dynamics.  Among these are
nucleation and growth in first-order phase transitions,
nonequilibrium kinetics associated with phase transitions, domain
wall movement in ferroelectrics, dielectric dissipation in
insulators and many others.  One such problem is presented by the
phenomenon of pressure amorphization.  This has been successfully
addressed at the conceptional level by augmenting a highly
simplified model Hamiltonian with Langivan dynamics,$^{6,7}$
showing how pressure amorphization can occur by the emergence of
metastable displacive disorder.  In the present paper, Zwanzig's
theory of nonlinear generalized Langevin equations$^{8}$ is used
to create a classical dynamic extension of the model Hamiltonian
method suitable for addressing such problems beyond the present
scope of that method.

In Langevin dynamics, a system evolves under the influence of its
internal dynamics; of a relaxation term or, more generally, of a
memory kernel which captures the coherent part of its interaction
with a thermal bath, and of the remaining stochastic forces from
the bath. In Section 2, we show how to decompose the
Born-Oppenheimer Hamiltonian of a material into a model system, a
bath, and an interaction between them in the form with which
Zwanzig starts. In Section 3, we recapitulate Zwanzig's
development to add clarity and to make the present paper more
nearly selfcontained. In Section 4, we transform Zwanzig's
abstract equations into explicit Langevin equations of motion for
the model coordinates and momenta.  We then comment on
implementation in Section 5 and conclude there that exploring the
computational feasibility of the scheme would be worthwhile. We
emphasize in Section 6 that the equilibrium properties of the
system are governed by a model Hamiltonian which differs from that
previously obtained by a term which captures the coherent
consequences of the interaction of the model degrees of freedom
with those omitted from the model Hamiltonians and comment on 
generalizing the formalism.
\\

\noindent{\bf {2.  THE SYSTEM PLUS BATH DECOMPOSITION}}\\

We make the Born-Oppenheimer approximation and confine ourselves
to classical dynamics for the nuclear coordinates.  The resulting
Hamiltonian ${\cal H}$ contains the nuclear coordinates and
momenta.  We divide the nuclear coordinates and momenta into two
groups, the system variables $\it{x}$ which contribute all the
anharmonicity in $\cal{H}$ and the bath variables $\it{y}$ the
dependence of $\cal{H}$ on which is harmonic. We are thus
supposing that the $\it{x}$ are the important structural
variables in analogy with the model Hamiltonian method, but are
retaining the remaining dynamic variables $\it{y}$ of the material
viewed as providing a thermal reservoir to the system variables.
$x$ and $y$ are column victors with coordinate subvectors Q and q,
respectively, and momentum subvectors P and p, respectively:
\begin{equation}
x=\left(\begin{array}{c}
Q\\
P
\end{array}\right)\ ,\ \ \ \ \ \ \
y=\left(\begin{array}{c}
q\\
p
\end{array}\right)
\end{equation}
${\cal H}$ can now be decomposed into a system part ${\cal
H}^{\prime}_{S}$(x), a bath part ${\cal H}_{B}$(y), and a
system-bath interaction ${\cal H}^{\prime}_{SB}(x,y)$,
\begin{equation}
{\cal H} = {\cal H}^{\prime}_{S}(x) + {\cal H}^{\prime}_{Sb}(x,y) + {\cal H}^{\prime}_{B}(y)
\end{equation}
The Q correspond to the local modes of the model Hamiltonian
method, and the potential energy part of ${\cal
H}^{\prime}_{S}$(x) corresponds to the entire model Hamiltonian
of the prior schemes$^{3-5}.$

The bath Hamiltonian ${\cal H}$$^{\prime}_{B}$(y) is harmonic,
\begin{equation}
{\cal H}^{\prime}_{B}(y) = \frac{1}{2} y^{T}\cdot K\cdot y,
\end{equation}
where y$^{T}$ is the row vector $(q^{T},p^{T})$.  K is a real,
symmetric,  positive-definite matrix in the y space with the
submatrices
\begin{equation}
K = \left(\begin{array}{c}
K_{qq}\\
K_{pq}
\end{array} \ \ \ \ \ \ \
\begin{array}{c}
K_{qp}\\
K_{pp}
\end{array}\right)
\end{equation}

In all cases contemplated here,  there are no velocity-dependent
forces so that $K_{qp}$ = K$_{pq}$ = 0.  The system - bath
interaction is linear in y but nonlinear in x in general,
\begin{equation}
{\cal H}^{'}_{SB} (x,y) = -\frac{1}{2} \left(a^{T}(x).y + y^{T}.
a(x)\right),
\end{equation}
where a(x) is a column vector in the y space which is in general a
nonlinear function of x.

To make contact with Zwanzig's development, ${\cal H}$ must be
rewritten.  ${\cal H}^{\prime}_{S}$ (x) contains both kinetic,
T$_{s}$(P), and potential, V$^{\prime}_{s}$(Q), energies,
\begin{equation}
{\cal H}^{\prime}_{S}(x) = T_{S}(P) + V^{\prime}_{S}(Q),
\end{equation}
with V$^{\prime}_{S}$ (Q) anharmonic.  We now rewrite ${\cal H}$ in the form used by Zwanzig
\begin{equation}
{\cal H}(x,y) = {\cal H}_{S}(x) + {\cal H}_{B}(x,y)
\end{equation}
so that
\begin{equation}
{\cal H}_{B}(x,y) = \frac{1}{2} \left( y-a(x) \right)^{T} \cdot K
\cdot \left( y-a(x)\right),
\end{equation}
absorbing ${\cal H}^{\prime}_{SB}$ into ${\cal H}_{B}$.  This
introduces an additional potential energy term into ${\cal
H}^{\prime}_{S}$ so that
\begin{equation}
{\cal H}_{S} (X) = T_{S}(P) + V_{S}(Q),
\end{equation}
\begin{equation}
V_{S}(Q) = V^{\prime}_{S}(Q) - \frac{1}{2}a^{T}(x) \cdot K \cdot a(x).
\end{equation}

Thus a(x) becomes the instantaneous equilibrium value of y, and
the non-negative term subtracted from $V^{\prime}_{S}$ (x) in
(9b) softens $V_{S}$(x).  Thus the potential energy of the model
system differs importantly from the original model Hamiltonian
and should yield improved equilibrium properties.  The
Hamiltonian (7) is the starting point of Zwanzig's development.\\

\noindent{\bf{3.  ZWANZIG'S FORMAL LANGEVIN DYNAMICS}}\\

\noindent{\bf{3.1 Kerner Dynamics}}\\

One can write Hamilton's equations of motion for the system variables in vector form
\begin{equation}
\left(\begin{array}{c}
\dot{Q}\\
\dot{P}
\end{array}\right) \ \  =
\left(\begin{array}{c}
{\partial \cal H}/{\partial Q}\\ -
{\partial \cal H}/{\partial P}
\end{array}\right)
\end{equation}

Kerner's procedure$^{9}$ collects these into a single equation
through the introduction of the antisymmetric matrix A operating
in the x space
\begin{equation}
A = \left(\begin{array}{c}
0\\
-1
\end{array} \ \ \ \ \ \ \
\begin{array}{c}
1\\
0
\end{array}\right).
\end{equation}

The equation of motion then becomes
\begin{equation}
\dot{x} = A \cdot \nabla_{x} {\cal H},
\end{equation}
\begin{equation}
\dot{y} = B \cdot \nabla_{y} {\cal H},
\end{equation}
where B is the analog of A in the y space.\\

\noindent{\bf 3.2  Motion in the y Space}\\

Introducing the subscript $\it{t}$ to specify the time at which
the dependent variables are evaluated, (14) becomes
\begin{equation}
\dot{y}_{t} = B \cdot K \cdot [y_{t} -a(x_{t})].
\end{equation}
the formal solution to (15) can be written as
\begin{eqnarray}
y_{t}& - & a(x_{t}) = e^{t B \cdot K}[y_{o}- a(x_{o})] \nonumber
\\ & - & \int^{t}_{o} dt^{\prime}e^{t{\prime} B \cdot K} \cdot
\nabla_{x} a(x_{t-t^{\prime}}) \cdot \cdot{x}_{t-t^{\prime}}
\end{eqnarray}
after partial integration subsequent to the direct integration of (15).\\

\noindent{\bf 3.3 Motion in the x-Space}\\

The equation of motion (13) for x is, more explicity,
\begin{equation}
\dot x_{t} = A \cdot \nabla_{x} {\cal H}_{S} - A \cdot
\nabla_{x}a^{T}(x_{t}) \cdot K \cdot [y_{t} - a(x_{t})].
\end{equation}
Substituting (16) into (17) yields a closed equation of motion
for $x$ which will turn out to be the desired Langevin equation,
\begin{eqnarray}
\dot {x}_{t} & = & A \cdot
\nabla_{x} {\cal H}_{S}(x_{t}) \nonumber \\
& + & \int^{t}_{o} dt^{\prime} A \cdot \nabla_x a (x_{t})^{T} \cdot
K \cdot e^{t^{\prime} B \cdot K}
\nabla_{x} a(x_{t-t^{\prime}}) \cdot \dot{x}_{t-t} \nonumber \\
& - & A \cdot \nabla_{x} a(x_{t}) \cdot K \cdot e^{t B \cdot K} [y_{o} -
a (x_{o})].
\end{eqnarray}
The first line on the right side of (18) is a deterministic,
conservative, nonlinear force.  The second line contains a
state-dependent, that is nonlinear, memory function which gives
rise to a dissipative force which is nonlocal in time.  The third
line contains a state-dependent response to a stochastic force.
In the next section we show the latter to be Gaussian random
colored noise.
\\

\noindent{\bf 3.4  The Random Force}\\

Define the vector $F_{t}$ in $y$ space as a force
\begin{equation}
F_{t} = -K \cdot e^{t B \cdot K} [y_{o} - a(x_{o})].
\end{equation}
The equation (18) for $\dot x_{t}$ is solved for $x_{t}$ given
$x_{o}$, the initial condition.  One could in principle also fix
$y_{o}$ as an initial condition and solve Eq. (18)
deterministically by molecular dynamics.  One would then, for
each value of $x_{o}$ chosen, have to compute the complete
trajectories to establish the statistical dynamics of the system
over and over for different $y_{o}$.  In doing so, any external
coupling to a reservoir forcing the system to relax towards
equilibrium is ignored.  Zwanzig instead makes the assumption
that the bath is in thermodynamic equilibrium at $t_{o}$ given
$x_{o}$.  This implies that the initial state $y_{o}$ has a
probability distribution
\begin{equation}
PROB[y_{o}|x_{o}] = \frac{e^{-\beta {\cal H}_{B}(y_{o},
x_{o})}}{\int d y^{\prime}_{o} e^{-{\cal H}_{B}
(y^{\prime}_{o},x^{\prime}_{o})}}
\end{equation}
so that $y_{o}$ is a Gaussian random variable (GRV).  F$_{t}$,
[19], is a superposition of GRV's and is therefore a Gausian random
variable itself with
\begin{eqnarray}
\langle F_{t} \rangle & = & 0, \\ 
\langle F_{t}F_{t^{\prime}} \rangle & = & k_{B} T  K \cdot e^{(t-t^{\prime}) B \cdot K}.
\end{eqnarray}\\

\noindent{\bf{3.5  The Langevin Equation}}\\

We can now interpret [16] as a generalized, nonlinear Langevin equation,
\begin{eqnarray}
\dot x_{t} = A \cdot [\nabla_{x} {\cal H}_{S}
(x_{t})] + \int^{t}_{o} dt^{\prime}
M(t,t^{\prime})\cdot {\dot x}_{t}^{\prime} \\ \nonumber
+ \nabla_{x} a^{T} (x_{t}) \cdot
F_{t}, \\ \nonumber
\end{eqnarray}
where
\begin{equation}
M(t_{,}t^{\prime}) = \nabla_{x}a(x_{t})^{T} \cdot K \cdot
e^{(t-t^{\prime})B \cdot K} \nabla_{x} a(x_{t^{\prime}}).
\end{equation}

\noindent{\bf{4.  EQUATIONS OF MOTION FOR Q AND P}}\\

\noindent{\bf{4.1  $\dot { \bf Q}$ and $\dot {\bf P}$}}\\ 

As asserted earlier, we exclude velocity-dependent forces,  
which simplifies the formalism.  We also absorb the nuclear masses into the
definition of the momenta so that
\begin{equation}
{\cal H}_{S}  =  \frac{1}{2} P \cdot P + V(Q),
\end{equation}
\begin{equation}
{\cal H}_{B}  =  \frac{1}{2} p \cdot p + (q - a_{q}(Q))^{T} \cdot K_{qq} \cdot (q-a_{q}(Q)), \\
\end{equation}
\begin{equation}
a(x)  =  
\left(\begin{array}{c}
a_{q}(Q)\\
0
\end{array}\right)\\
\end{equation}
\begin{equation}
\nabla_{x} a(x) = \left(\begin{array}{c}
\nabla_{Q} a_{q}(Q)\\
0
\end{array} \ \ \ \ \ \ \
\begin{array}{c}
0\\
0
\end{array}\right).
\end{equation}
If we define
\begin{equation}
L(t) = K \cdot e^{t B \cdot K},
\end{equation}
then $L_{qq}$ is its only nonzero submatrix so that
\begin{equation}
M(t,t^{\prime}) = \left(\begin{array}{c}
(\nabla_{Q}a_{q}\left(Q_{t}\right))^T
\cdot L_{qq}(t-t^{\prime}) \cdot \nabla_{Q} a_{q}(Q_{t.})\\
0
\end{array} \ \ \ \ \ \ \
\begin{array}{c}
0\\
0
\end{array}\right).
\end{equation}

All this results in the much simpler, explicit equations of motion
\begin{eqnarray}
\dot Q_{t} & = & P_{t},\\
\dot P_{t} & = & -\nabla_{Q}V(Q_{t}) -\int^{t}_{o} dt^{\prime}
M_{pp}(t,t^{\prime}) \cdot P_{t^{\prime}} \nonumber
\\
& - & \nabla_{Q}a_{q}(Q_{t})^T \cdot F_{q},\\
M_{pp}(t,t^{\prime}) & = & \nabla _{Q}a_{q}(Q_{t})^T \cdot
L_{qq}(t-t^{\prime})\cdot \nabla_{Q}a_{q}(Q_{t}),\\
F_{qt} & = &  - L_{qq}(t) \cdot [q_{o} - a_{q} (Q_{o})],\\
\langle F_{qt}F_{qt^{\prime}} \rangle & = & k_{B}TL_{qq}({t-t^{\prime}}).
\end{eqnarray}
The remaining task is to simplify $L_{qq}(t-t^{\prime})$.
\\

\noindent {\bf 4.2 Bath Normal Modes}\\

We now transform to the normal modes of K in the y space.  K and
B are now comprised entirely of a set of 2$\times$2 diagonal
submatrices $K_{\lambda\lambda}$ and B$_{\lambda\lambda}$ along
the main diagonal with
\begin{eqnarray}
K_{\lambda\lambda} & = & \left(\begin{array}{c}
\omega^{2}_{\lambda}\\
0
\end{array} \ \ \ \ \ \ \
\begin{array}{c}
0\\
0
\end{array}\right).
\end{eqnarray}
and correspondingly
\begin{eqnarray}
B_{\lambda\lambda} & = & \left(\begin{array}{c}
0\\
1
\end{array} \ \ \ \ \ \ \
\begin{array}{c}
-1\\
0
\end{array}\right).
\end{eqnarray}

Inserting these into $L_{qq}(t-t^{\prime})$ yields
\begin{eqnarray}
L_{q \lambda,q \lambda^{\prime}}(t-t^{\prime}) = \omega_{\lambda}
cos \omega_{\lambda} (t-t^{\prime}) \delta_{\lambda \lambda{^\prime}},
\end{eqnarray}
greatly simplifying (35) and leading to
\begin{equation}
M_{PP}(t,t^{\prime}) = \int d\omega g(\omega) \omega^{2} cos \omega (t-t^{\prime}) 
W_{PP}(\omega; t,t^{\prime}),
\end{equation}
\begin{equation}
W_{PP}(\omega;t,t^{\prime}) =
[\sum_{\lambda}\delta(\omega_{\lambda}-\omega) \nabla_{Q}a_{q\lambda} 
(Q_{t}) \nabla_{Q}a_{q\lambda} (Q_{t^{\prime}})] /g(\omega),
\end{equation}
\begin{equation}
g(\omega) = \sum_{\lambda} \delta (\omega_{\lambda}-\omega)
\end{equation}

From (40), g($\omega$) is the total phonon density of states of the bath.

Still more explicitly,  let $\ell$ be the index of the degrees of
freedom of S.  The equations of motion become
\begin{eqnarray}
\dot Q_{\ell t} & = & P_{\ell t},\\
\dot P_{\ell t} & = &  -\nabla_{Q \ell} V(Q) - \sum_{\ell}
\int^{t}_{o} dt^{\prime} M_{\ell\ell}^{\prime}(t,t^{\prime})
P_{\ell} t^{\prime},\\
M_{\ell \ell^{\prime}} & = & \int d\omega g(\omega) \omega^2 cos \omega (t -t^{\prime})
W_{\ell \ell^{\prime}} (\omega; t,t^{\prime}),\\
W_{\ell \ell^{\prime}} (\omega; t,t^{\prime}) & = & [\sum_{\lambda} \delta(\omega_{\lambda}-\omega)
\nabla_{Q_\ell} a_{q_\lambda}(Q)\nabla_{Q_\ell^\prime} a_{q_\lambda}(Q)]/g(\omega),\\
\langle F_{q_\lambda t} F_{q_\lambda^\prime t^\prime} \rangle & = & k_B T \omega_\lambda^2
cos \omega_\lambda (t - t^\prime) \delta_{\lambda \lambda^\prime}.
\end{eqnarray}

\noindent{\bf{5.  COMMENTS ON IMPLEMENTATION}}\\

To focus our attention,  we discuss as an illustrative example the
dynamics of a first-order structural phase transition under
pressure from crystalline phase C to crystalline phase
C$^{\prime}$. We suppose that C is the simpler, more symmetric
structure.  The transition is associated with the development of
a soft branch of the phonon spectrum of C.  There is a spinodal
in the pressure-temperature plane at which C becomes locally
unstable. Implementation of the Langevin dynamics for studying
the phase transition proceeds in a series of steps which we
sketch in the
following.\\

\noindent{\bf{5.1  Decomposition into System Plus Bath}}\\

Consider pressures such that C is unstable at T=0.  Calculate the
harmonic normal mode spectrum of C.  Establish which are the
unstable modes or branches of C.  Construct local modes from the
unstable modes and identify their amplitudes with Q.  All other
normal modes are allocated to the bath, and K and the normal
modes $q_{\lambda}$ are determined.
\\

\noindent{\bf{5.2  Mapping from ${\cal H}$ to ${\cal H}_{S}$ + ${\cal H_{B}}$}}\\

One first assumes simple analytic forms for the Q dependence of
V(Q) and a(Q) motivated by the insights underlying the
decomposition of the material into system plus both. Practical
considerations limit both the complexity of the Q dependence of
the forms assumed for V(Q) and a(Q) and the fineness with which
the Q and q spaces are sampled.  V(Q) is then obtained from
first-principles electronic-structure computations of the total
energies for the fixed sample of Q values which are
{\it fully relaxed} with regard to the sampled q 
variables. The $a$(Q) are then simply equal to the
relaxed values of the q$_{\lambda}$, according to Eqs. (2)-(9).
The parameters of the simple analytic forms for V(Q) and a(Q) are
then determined by a best fit to the numerical data.  Thus the
computational cost of generating the necessary information on
which to base the Langevin simulations is increased beyond
that of the previous model-Hamiltonian constructions only by
the relaxation.\\

\noindent {\bf5.3  Constructing the Langevin Equation}\\

Knowledge of V(Q) yields the internal forces $\nabla_{Q_\ell}V(Q)$. 
Knowledge of $a(Q)$ yields the state-dependent quantities
$\nabla _{Q_\ell}a_{\lambda}(Q)$.  Knowledge of K yields the
frequency spectrum g($\omega$), the random force autocorrelation
function, and, together with $\nabla _{Q_\ell} a_{\lambda} (Q)$,
W$_{\ell \ell^{\prime}} (t-t^{\prime})$.  Thus everything
entering the Langevin equations is known once the parameters of
${\cal H_{S}}$ +
${\cal H_{B}}$ are established by the mapping of Section 5.2.\\

\noindent {\bf 5.4 Solving The Langevin Equation}\\

Apart from the issue of the number of variables to keep track of,
there are three essential difficulties in standing in the way of
implementing numerically the solution of the Langevin equations:
1)  Each of the two terms in the equation for $P_{\ell}$ is
nonlinear.  2)  There is a memory, and the noise is colored, which
require remembering Q, P, and F$_{q \lambda}$ over an interval of
time longer than a relevant bath phonon period.  3)  There are two
quite different time scales in the problem, the bath phonon period
and the characteristic relaxation times implicit in the memory
function.  Issues 1.) and 3.) are encountered in molecular
dynamics.  Issue 1.) is not serious, and issue 3.) has been
confronted with varying degrees of success in MD.  Thus the main
difficulty is the increase in memory requirement associated with
issue 2.). Assuming that issue 2.) can be overcome, one would
start the simulation of the first order transition from C to
C$^{\prime}$ by drawing an initial value of x from the truncated 
canonical ensemble
\begin{equation}
\rho(x) = \frac{e^{-\beta {\cal H}_{S}(x)}}{\int dx^{\prime} 
e^{-\beta {\cal H}_S (x^{\prime})}}
\end{equation}
with x and x$^{\prime}$ limited to the basin of attraction of C in
V$_{S}$(Q).  One would then choose a time step which permitted
adequate sampling of the memory function and the random force
autocorrelation function.  Forward time integration could proceed
by a suitable generalization of the Verlet algorithm adapted for
different-integral equations.  Updating the random force at each
time step need not be onerous. In molecular dynamics only P and Q
need be remembered at each time step, and only $\nabla _{Q}V$
needs to be updated.  In the current Langevin dynamics, P, Q
$\nabla _{Q}V, \nabla _{Q}a(Q)$, and F$_{q}$ all need to be
updated. Thus, the upper bound to the number of degrees of freedom
which can feasibly be incorporated into ${\cal H_{S}}$ will turn
but to be much less than can be incorporated into a standard
molecular dynamics computation with classical pairwise forces.
The latter now exceeds 10$^{6}$.  Losing as many as three to four
orders of magnitude in feasible cell size would still allow quite
complex problems to be tackled.  We therefore propose that
investigating the computational complexity of the proposed
Langevin dynamics scheme with the goal of establishing its
feasibility and, if feasible, its size limitations would be
highly worthwhile.
\\

\noindent {\bf 6.  CONCLUSIONS}\\

The current procedure for generating the ``model'' or
``effective''  Hamiltonian from first-principles
calculations$^{3-5}$, generates only the potential
$V^{\prime}_{S}$(Q) in ${\cal H^{\prime}_{S}}$(x), Eq.(6).  By
transforming the full Hamiltonian of Eq.(2) into the Zwanzig
form Eq.(7), we have added an additional, softening term to
$V^{\prime}_{S}$(Q) to arrive at a system potential $V_{S}(Q)$,
Eq.(10), which includes all of the coherent effects of the
interaction between the reduced model system and the rest of the
material.  If one constructs the master equation for the
probability distribution of x at t given x$_{o}$ at t = 0,
PROB[x,t | x$_{o}$], a generalized Fokker-Planck equation, one
finds that PROB[x,t | x$_{o}$] asymptotically approaches a unique
stationary solution which is a canonical ensemble for the
Hamiltonian ${\cal H}_{S} (x)$, not ${\cal H}^{\prime}_{S}(x)$,
\begin{equation}
PROB[x,t|x_{o}] \rightarrow {e^{-\beta {\cal H}_s(x)} \over
{\int dx^\prime e^{-\beta {\cal H}_s(x^\prime)}}},t\rightarrow\infty,
\end{equation}
which differs from $\rho$(x) in Eq.(47) because there is no
truncation here.  This change is a direct consequence of the form
of ${\cal H_{B}}$, Eq.(8), which states that the bath variables
relax towards and oscillate around their instantaneous
equilibrium positions $a(x)$.  Thus, the use of $V^{\prime}_{S}$(Q)
and ${\cal H}^{\prime}_{S}$(x) for the calculation of equilibrium
properties is conceptually wrong.

A quantitative analysis of the errors in the current
first-principles effective-hamiltonian approach to
ferroelectric perovskites has recently been carried out by
Tinte {\it et al}$^{15}$.  They show that not calculating the fully
relaxed $V_{S}$(Q) and calculating instead the unrelaxed
$V^{\prime_{S}}$(Q) does not, for the materials studied,
introduce a major error.  Nevertheless, as discussed in Section
5.2, the fully relaxed calculation is necessary to generate the
interaction amplitudes $a$(Q) essential for implementing the
Langevin formalism.  Tinte {\it et al.} report that the major error
arises from ignoring thermal expansion.  The import of that
conclusion for the present analysis is that anharmonicity in the
bath coordinates q should be included.  The only practical way to
do so without losing the distinction between system and bath is
through the use of a psendo-harmonic approximation$^{16}$ to the
time evolution and statistical distribution of the bath variables
$y$.  We reserve that generalization of the present formalism for
a future publication.

We reiterate the  conclusion of Section 5.4 that exploring the
feasibility of numerical implementation of the present Langevin
dynamics scheme would be highly worthwhile because of the great
extension of the reach of model Hamilton methods$^{3-5, 10-14}$
which could result.

This work was supported by the Center for Piezoelectric Design
under ONR Grant N00014-01-1-0365. The assistance of
J. B. Neaton is gratefully acknowledged.



\end{document}